          [2001/04/25 1.1 (PWD)]
\documentclass[a4paper,twocolumn]{esapub} 
\usepackage{times}
\usepackage{natbib}
\usepackage{graphicx}

\title{Asteroseismology of exoplanets-host stars: a link between the two scientific programmes of COROT}
\author{Sylvie Vauclair}
\author{Matthieu Castro}
\author{St\'ephane Charpinet}
\author{Marion Laymand}
\author{M\'elanie Soriano}
\author{G\'erard Vauclair}
\affil{Laboratoire d'Astrophysique; Observatoire Midi-Pyr\'en\'ees; Universit\'e Paul Sabatier, Toulouse, France}
\author{Michael Bazot}
\affil{Institut for Physik og Astronomi, Aarhus Universitet,Denmark}
\author{Francois Bouchy}
\affil{Institut d’Astrophysique de Paris, France}

\begin{document}


\maketitle

\begin{abstract}
Studying the internal structure of exoplanets-host stars compared to that of similar stars without detected planets is particularly important for the understanding of
 planetary formation. 
The observed overmetallicity of stars with planets may be a hint in that respect. Although it is obviously related to the physical processes which occur during the early phases of planetary formation, the origin of this overmetallicity is unclear. It may be either primordial or related to accretion processes or both.
In this framework, asteroseismic studies represent an excellent tool to 
determine the structural differences between stars with and without detected planets. The two different missions of COROT are linked in this programme: the detection of new planets and the seismic studies of their host stars share the same goal of a better understanding of planetary formation and evolution. The COROT main target HD52265, which is known to host at least one giant planet, will be observed continuously during five months: many interesting results are expected from this long run. Meanwhile, stellar oscillations will be searched for in all stars around which new planets will be discovered.
\end{abstract}

\section{Introduction}

Asteroseismology of exoplanets-host stars is an important new subject which will be largely included in COROT programmes. One of the COROT main targets, HD52265, which will be observed during the second long run in the anti-centre direction, is known to possess at least one giant planet. It will be followed during five consecutive months, which will provide the data needed to precisely study its internal structure. Meanwhile, all the stars with newly detected planets will be seismically studied to detect their oscillations. This additional programme will allow for statistics on these new planets-host stars and will eventually be used to choose a specific sample for future space or ground based observations.

Among the differences between stars with or without detected planets, 
the observed overmetallicity of the first ones compared to the second ones needs to be understood (Santos et al.\cite{santos03} and \cite{santos05}, Gonzalez \cite{gonzalez03},
 Fischer and Valenti \cite{fischer05}). Two scenarios are possible to account for this observed overmetallicity : the “primordial origin”, which assumes that the stars formed out of an overmetallic nebula, and the “accretion origin” for which the observed metallicity is due to accretion of hydrogen poor material onto the star during planetary formation.

Bazot and Vauclair \cite{bazot04} 
pointed out that the evolution of stars with masses around 1.1 M$_{\odot}$ is very sensitive to their internal metallicity, due to the possible formation of a convective core. In this particular region of the HR diagram, main-sequence stars with solar internal metallicities have no convective cores while overmetallic stars develop convection in their central regions: evolutionary tracks computed with different internal metallicities may cross the same point in the HR diagram while they correspond to 
models of different masses and different past histories. 

This behavior was used to try to derive whether the exoplanet host star $\mu$ Arae is overmetallic from its surface down to its center (overmetallic scenario) or only in its outer layer (accretion scenario). The star was observed during height nights with the HARPS spectrometer (on the 3m60 telescope in La Silla, Chile) in June 2004. Up to 43 p-modes could be identified (Bouchy et al. \cite{bouchy05}) and a detailed modelisation could 
be achieved (Bazot et al.\cite{bazot05}). Possible tests which could be used to determine the internal structure of the star and more specifically its internal metallicity were discussed. It was shown that the external parameters for the star (effective temperature, gravity, luminosity) could be obtained with a better precision than derived from spectroscopy.
Unfortunately the modes 
with frequencies around and above 2.5~mHz could not be identified with enough precision to reach a 
definitive answer about the internal metallicity of this star. More work is presently being done on this subject.

For the “COROT star” HD 52265, which will be observed during five consecutive months, we expect much more precise data, which will lead to better frequency determinations and mode identifications. We will use the same kind of tests with more efficiency and we expect to reach precise conclusions on the internal structure and past history of this exoplanet-host star. 

In the following two sections, we first recall the most relevant asteroseismic tests used for $\mu$ Arae, as they will act as examples for further studies on other stars; then we show preliminary results obtained in the modelisation of HD52265 as a preparation for COROT observations.

\section{Asteroseismic tests: the example of $\mu$ Arae }

The exoplanets-host star $\mu$ Arae (HD160691, HR6585, GJ691) is a G5V star with a visual magnitude V=5.1, an Hipparcos parallax $\pi$=65.5$\pm$0.8 mas, which gives a
 distance to the Sun of 15.3 pc and a luminosity of $\log L/L_{\odot} = $0.28$\pm$0.012. Spectroscopic observations by various authors gave five different effective 
temperatures and metallicities (see references in Bazot et al. \cite{bazot05}). The HARPS observations allowed to identify 43 oscillation modes of degrees $\ell$=0 to 
$\ell$=3 
(Bouchy et al. \cite{bouchy05}). From the analysis of the frequencies and comparison with models, the values T$_{eff}$=5813$\pm$40 K and [Fe/H]=0.32$\pm$0.05 dex were 
derived: 
these values, which lie inside the spectroscopic boxes, are obtained with a much better precision than spectroscopy.

\subsection{The stellar models }

Models were computed using the TGEC (Toulouse-Geneva stellar evolution code), with the OPAL equation of state and opacities (Rogers \& Nayfonov \cite{rogers02},
 Iglesias \& Rogers \cite{iglesias96}) and the NACRE nuclear reaction rates (Angulo et al. \cite{angulo99}). In all models microscopic diffusion was included using 
the Paquette prescription (Paquette et al. \cite{paquette86}). The treatment of convection was done in the framework of the mixing length theory and the mixing length 
parameter was adjusted as in the Sun ($\alpha = 1.8$). Three kinds of models were computed, according to the initial assumptions: overmetallic models with two different initial helium values, and accretion models. 
For the computations of overmetallic models, the helium value is crucial as differences in helium may lead to completely different evolutionary tracks. This question is related to the stellar and planetary formation: if the stellar system formed inside an overmetallic interstellar cloud, was this cloud also helium-enriched as obtained in the chemical evolution of galaxies or not? This depends on the stellar mass function and there is no clear answer at the present time. 

For the computations of the accretion models, the way accretion occurred, as well as the composition of the accreted matter is also a subject of debate. In the models, the most simple assumption was used: instantaneous accretion of matter with solar composition for metals and no light elements, at the beginning of the main-sequence. Neither extra-mixing nor overshoot were included.

\subsection{The oscillations }

Adiabatic oscillation frequencies were computed using an updated version of the code PULSE described in Brassard et al.\cite{brassard92}. For each evolutionary track, many models were computed inside the observed spectroscopic boxes in the HR diagram, but only those which could reproduce the observed echelle diagram were kept for subsequent tests (Figure 1). For these models the large separation $\Delta \nu_{l} = \nu_{n+1, l} - \nu_{n, l}$ is exactly 90 $\mu$Hz: a difference of only 0.5 $\mu$Hz completely destroys the fit with the observations. 

\begin{figure}
\centering
\includegraphics[angle=0,totalheight=5.5cm,width=8cm]{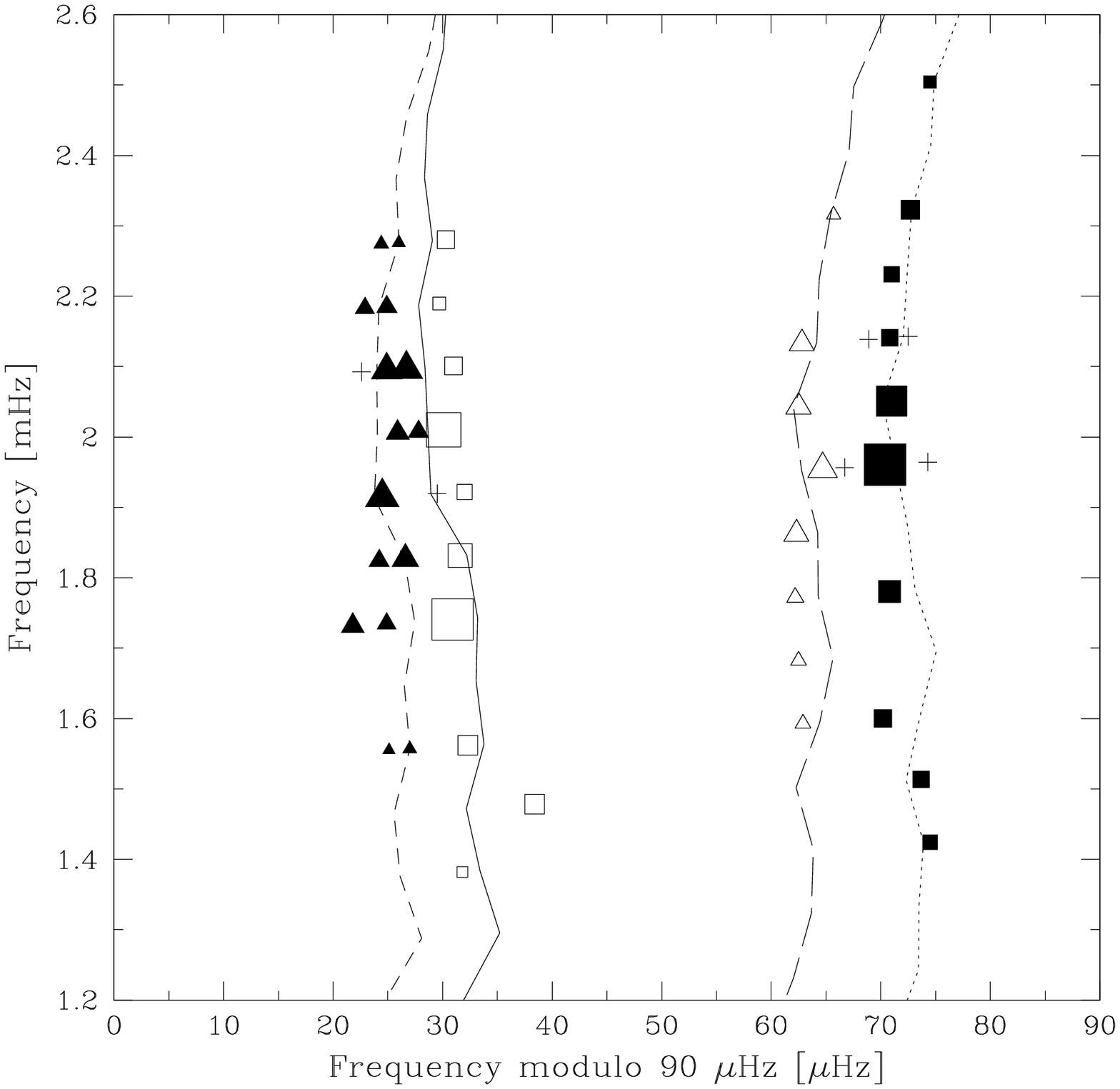}
\includegraphics[angle=0,totalheight=5.5cm,width=8cm]{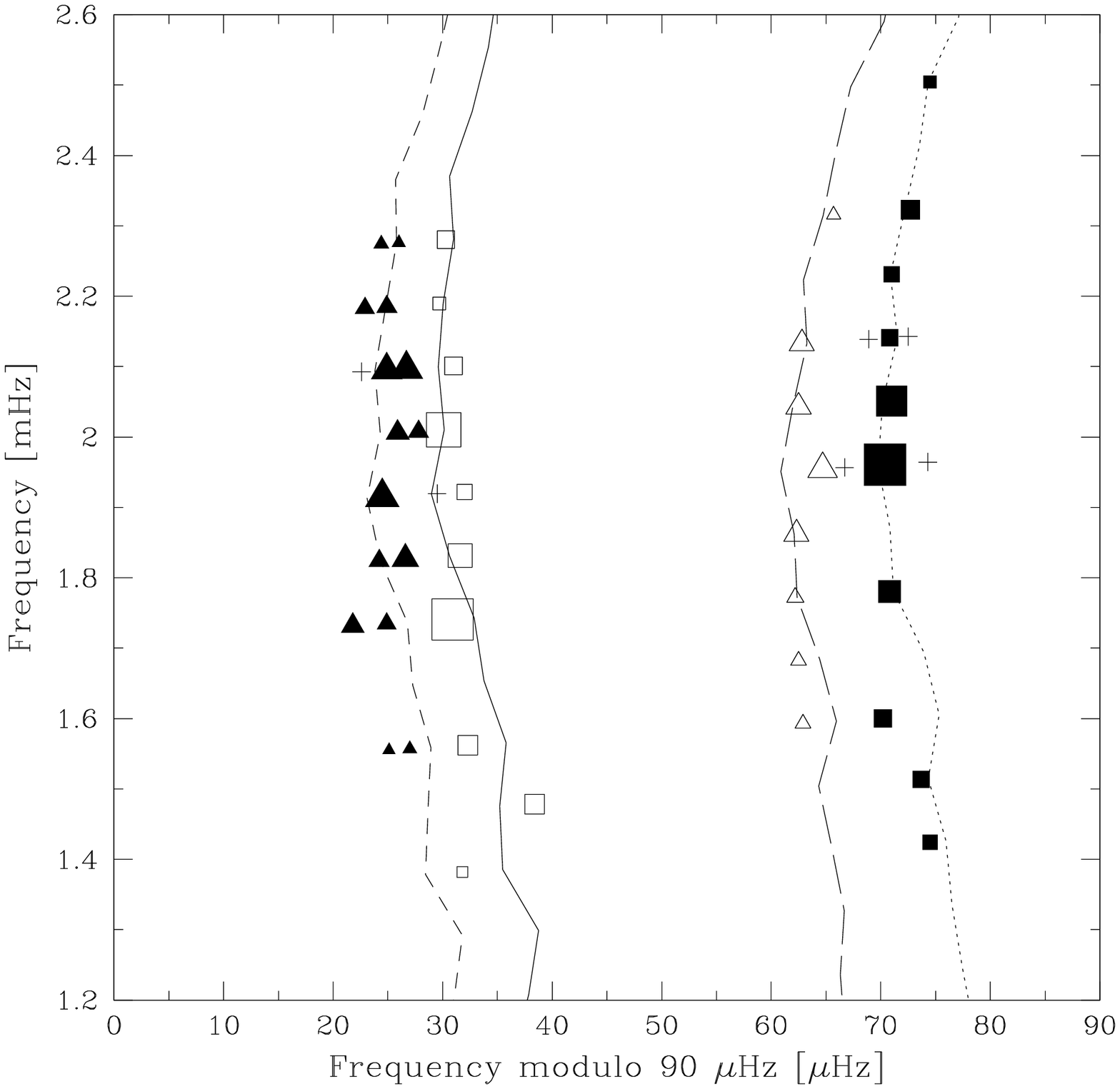}
\caption{ Echelle diagrams for two models of the star $\mu$ Arae: one with primordial overmetallicity (upper panel) and one with accretion (lower panel). The ordinate 
represents the 
frequencies of the modes and the abscissa the same frequencies modulo the large separation, here 90 $\mu$Hz. The lines represent the computations, respectively, from left to 
right : $\ell$=2, 0, 3 and 1 and the symbols represent the observations (after Bazot et al. \cite{bazot05}).\label{fig1}}
\end{figure}

Figure 2 displays the computed and observed small separations: $\delta \nu = \delta \nu_{n,l} - \nu_{n-1,l+2}$. As this quantity is very sensitive to the deep stellar 
interior (Tassoul \cite{tassoul80}, Roxburgh \& Vorontsov \cite{roxburgh94}), it was proposed as a good indicator of the presence of a convective core in stars around 1.1 M$_{\odot}$.
There is indeed an important difference between the overmetallic and the accretion cases for the $l=0,2$ small separations: in the overmetallic case they rapidly decrease at large frequencies and even become negative at some point while in the accretion case the curve is much flatter. This behavior also appears in the echelle diagrams: for the overmetallic case, the theoretical lines for $l=0$ and $l=2$ come closer for larger frequencies and even cross around $\nu = 2.7$ mHz. Here it seems that the observed points give a better fit with the accretion case than the overmetallic case but this has to be confirmed.

\begin{figure}
\centering
\includegraphics[angle=0,totalheight=5.5cm,width=8cm]{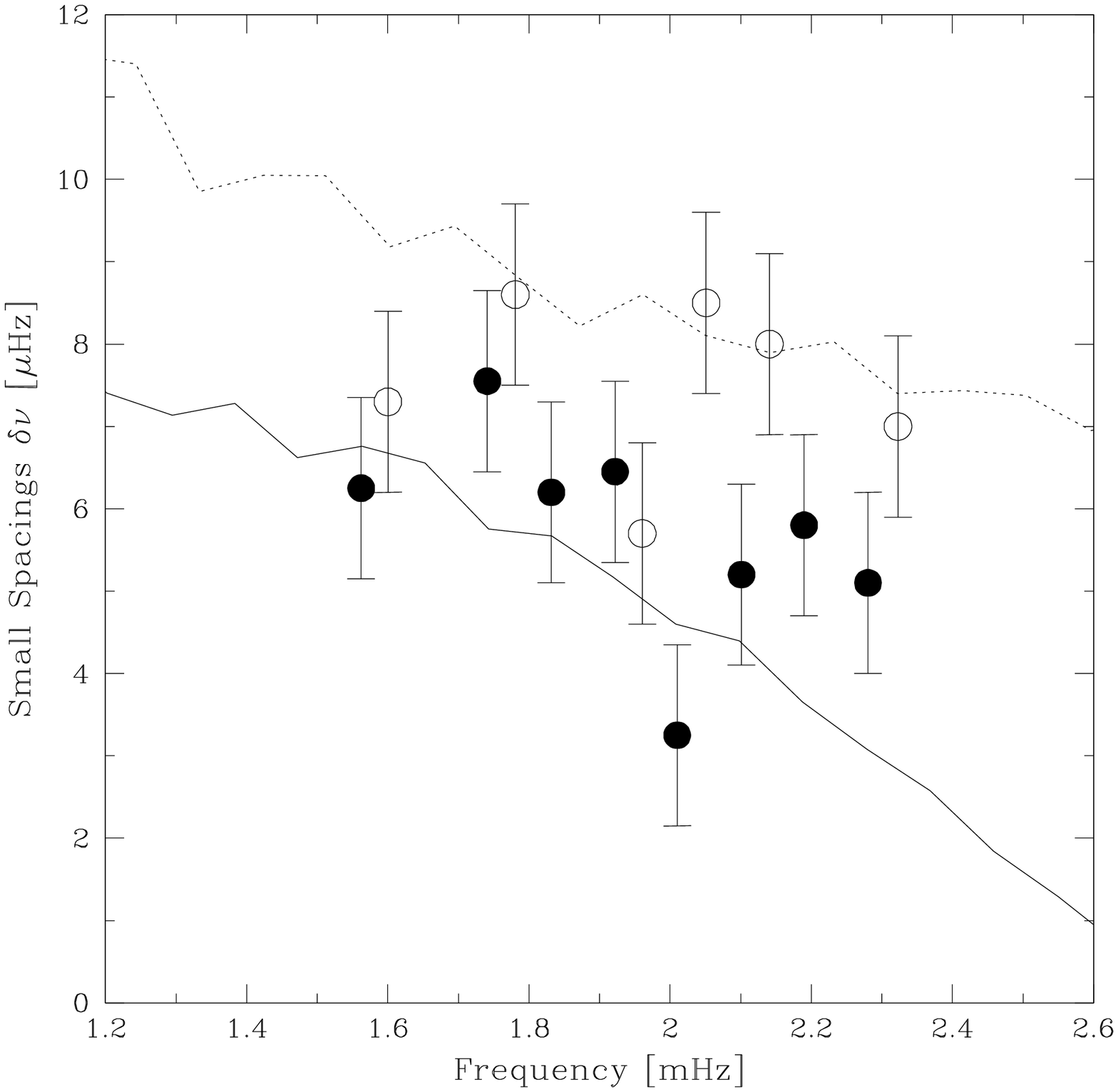}
\includegraphics[angle=0,totalheight=5.5cm,width=8cm]{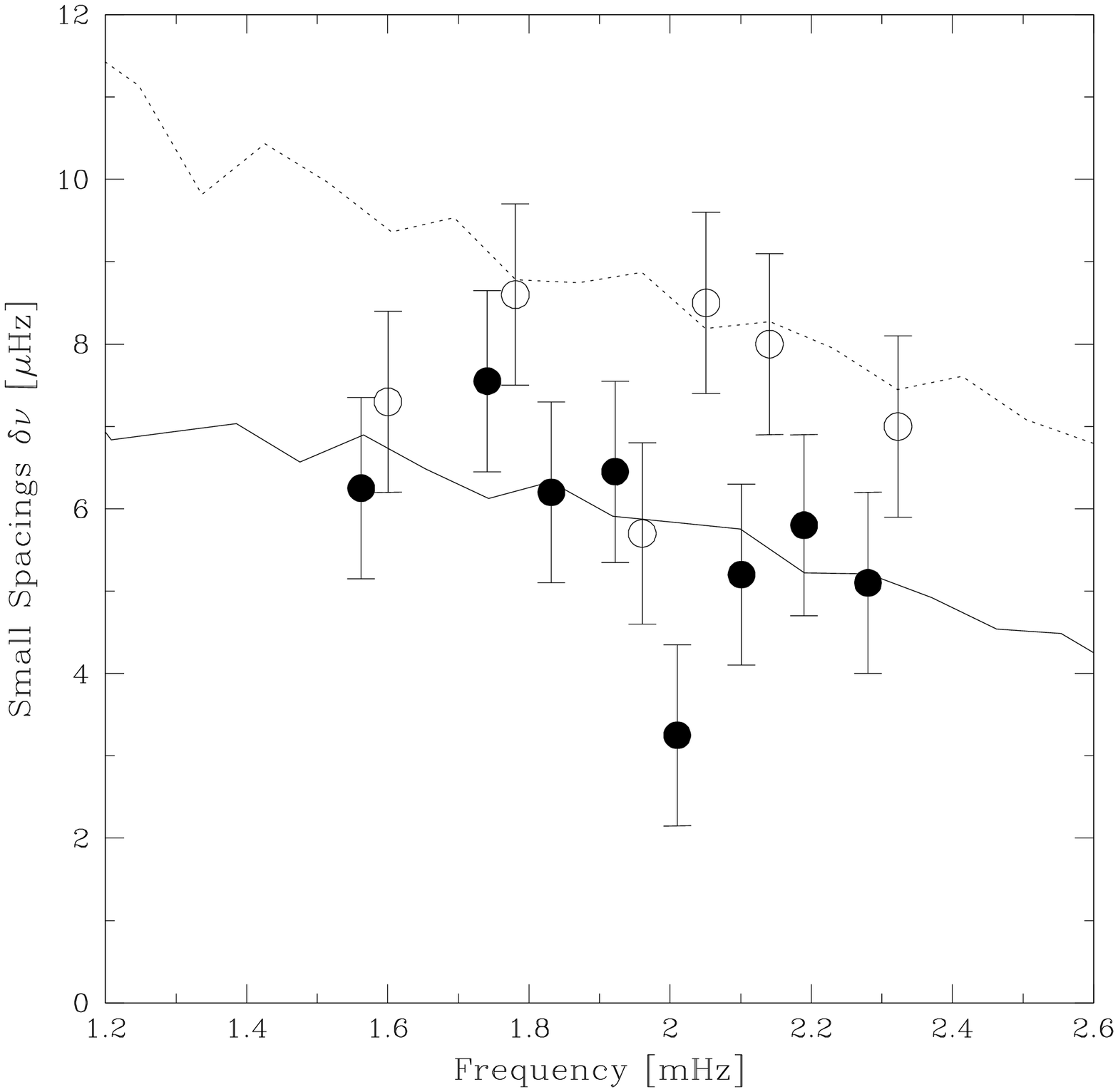}
\caption{ Small separations for the models presented in Figure 1; solid lines : l = 0,2 ; 
dotted lines: l = 1,3 (after Bazot et al. \cite{bazot05}).\label{fig2}}
\end{figure}

\section{Modelisation of the “COROT Star” HD 52265} 

Preliminary computations and modelisation of HD52265 have been done using the same techniques as for $\mu$ Arae, as a preparation of the future observations with COROT 
(Soriano, Laymand, Vauclair \& Vauclair, \cite{soriano06}, hereafter SLVV06). Evolutionary tracks and stellar models were computed with the TGEC (Toulouse Geneva Evolutionary Code) 
code and the adiabatic oscillation frequencies with the code PULSE.
Some examples of the results of this modelisation are given below.

\subsection{Spectroscopic Observations}

Five different groups of observers have given external parameters for HD52265. They are listed in Table 1. From the Hipparcos parallax, using Flower \cite{flower96}
 bolometric 
correction, we obtain a luminosity L/L$_{\odot}$=1.94$\pm$0.16 (SLVV06). These observations are used as constraints for the first step of model computations. As discussed
 in the next section, it is already possible to eliminate some of these values. Later on, we will go further with asteroseismology: COROT observations will lead to a much 
better precision on the external parameters of this star.

\begin{table}
\caption{Effective temperature, gravity and  metallicity of HD52265 as obtained from spectroscopic observations.
The references are: Santos et al. \cite{santos04} (SIM04); Gonzalez et al. \cite{gonzalez01} (GLTR01); Fischer \& Valenti \cite{fischer05} (FV05); 
Takeda et al. \cite{takeda05} (TOSKS05) and  Gillon \& Magain \cite{gillon06} (GM06)}
\label{tab1}
\begin{flushleft}
\begin{tabular}{cccccccc} \hline
\hline
T$_{\mbox{eff}}$ & log g & [Fe/H] & authors \cr
  \hline
 6103$\pm$52 & 4.28$\pm$0.12 & 0.23$\pm$0.05 & SIM04 \cr 
 6162$\pm$22 & 4.29$\pm$0.04 & 0.27$\pm$0.02 & GLTR01 \cr
 6076$\pm$44 & 4.26$\pm$0.06 & 0.19$\pm$0.03 & FV05 \cr
 6069$\pm$15 & 4.12$\pm$0.03 & 0.19$\pm$0.03 & TOSKS05 \cr
 6179$\pm$18 & 4.36$\pm$0.03 & 0.24$\pm$0.02 & GM06 \cr
\hline
\end{tabular}
\end{flushleft}
\end{table}

\subsection{Models and seismic tests}

We have computed evolutionary tracks for three different metallicities, as given from spectroscopic observations. For each metallicity, we have only kept the tracks crossing 
the corresponding error box(es) in the log g - log T$_{\mbox{eff}}$ diagram. Then we have computed oscillation frequencies for models which satisfied the constraints. 
Complete results will be given in a forthcoming paper (SLVV06). Here we show in Figure 3 two extreme examples, for metallicities [Fe/H] = 0.19 and 0.27.

\begin{figure}
\centering
\includegraphics[angle=0,totalheight=5.5cm,width=8cm]{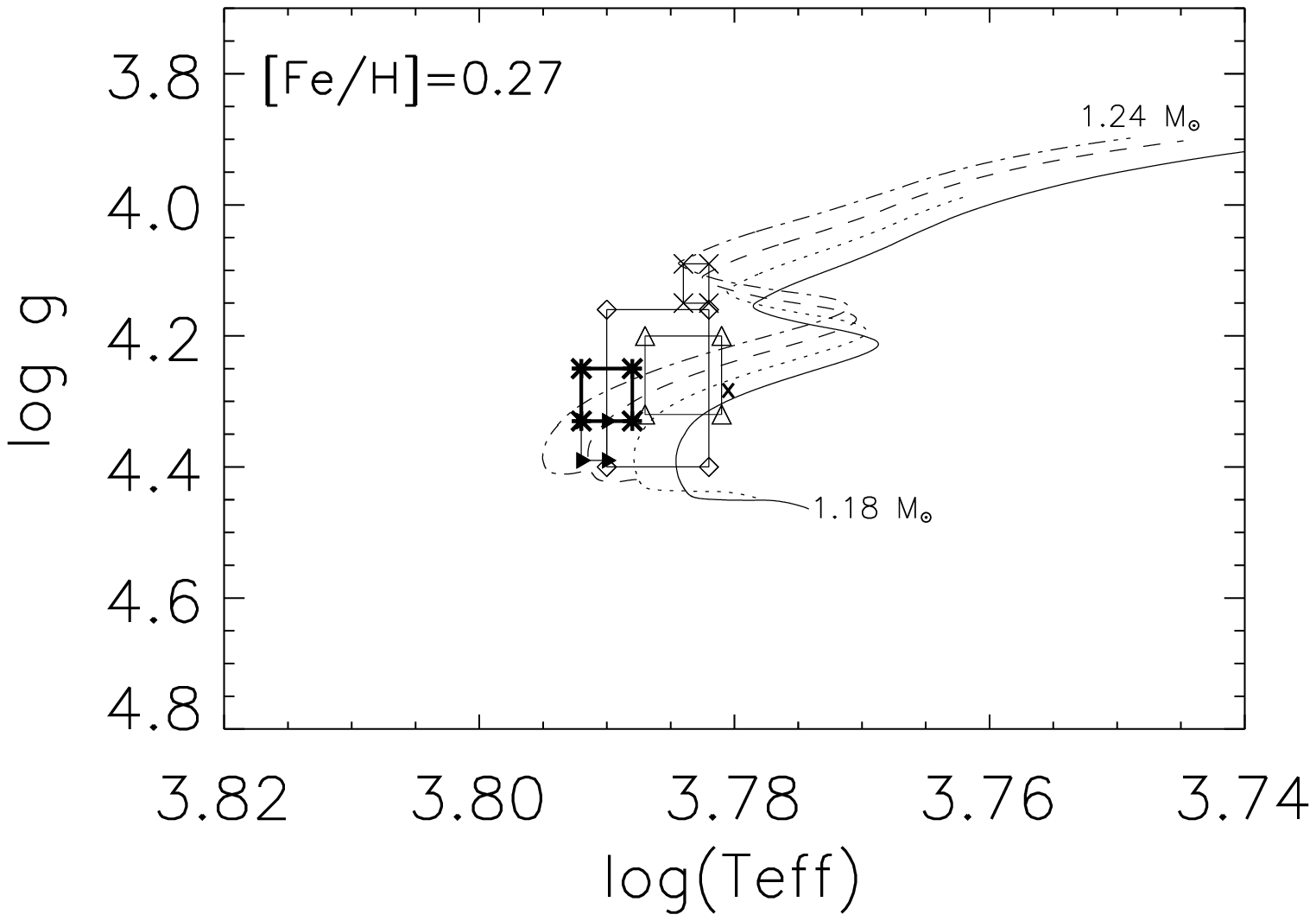}
\includegraphics[angle=0,totalheight=5.5cm,width=8cm]{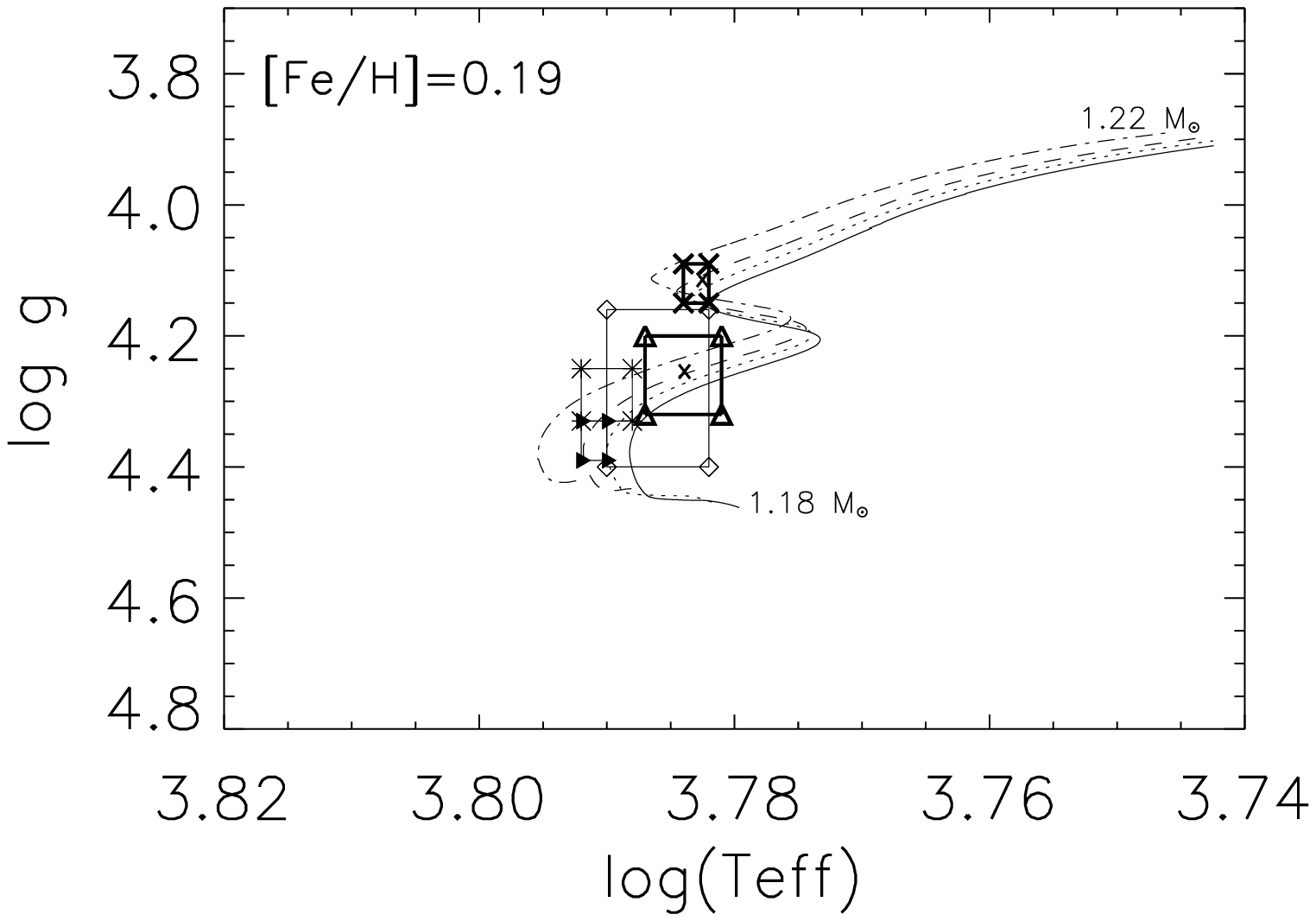}
\caption{ Evolutionary tracks computed in the log g - log T$_{\mbox{eff}}$ diagram for metallicities 0.27 (upper panel) and 0.19 (lower panel). Also shown are the five error boxes: SIM04 (diamonds), GLTR01 (stars), FV05 (white triangles) TOSKS05 (crosses), GM06 (black triangles). In the 0.19 case, another set of tracks computed for more massive stars 
(M $\approx$ 1.3 M$_{\odot}$) cross the Takeda et al. (TOSKS05) error box while at the end of the main sequence. These tracks are not presented here.\label{fig3}}
\end{figure}

\begin{figure}
\centering
\includegraphics[angle=0,totalheight=5.5cm,width=8cm]{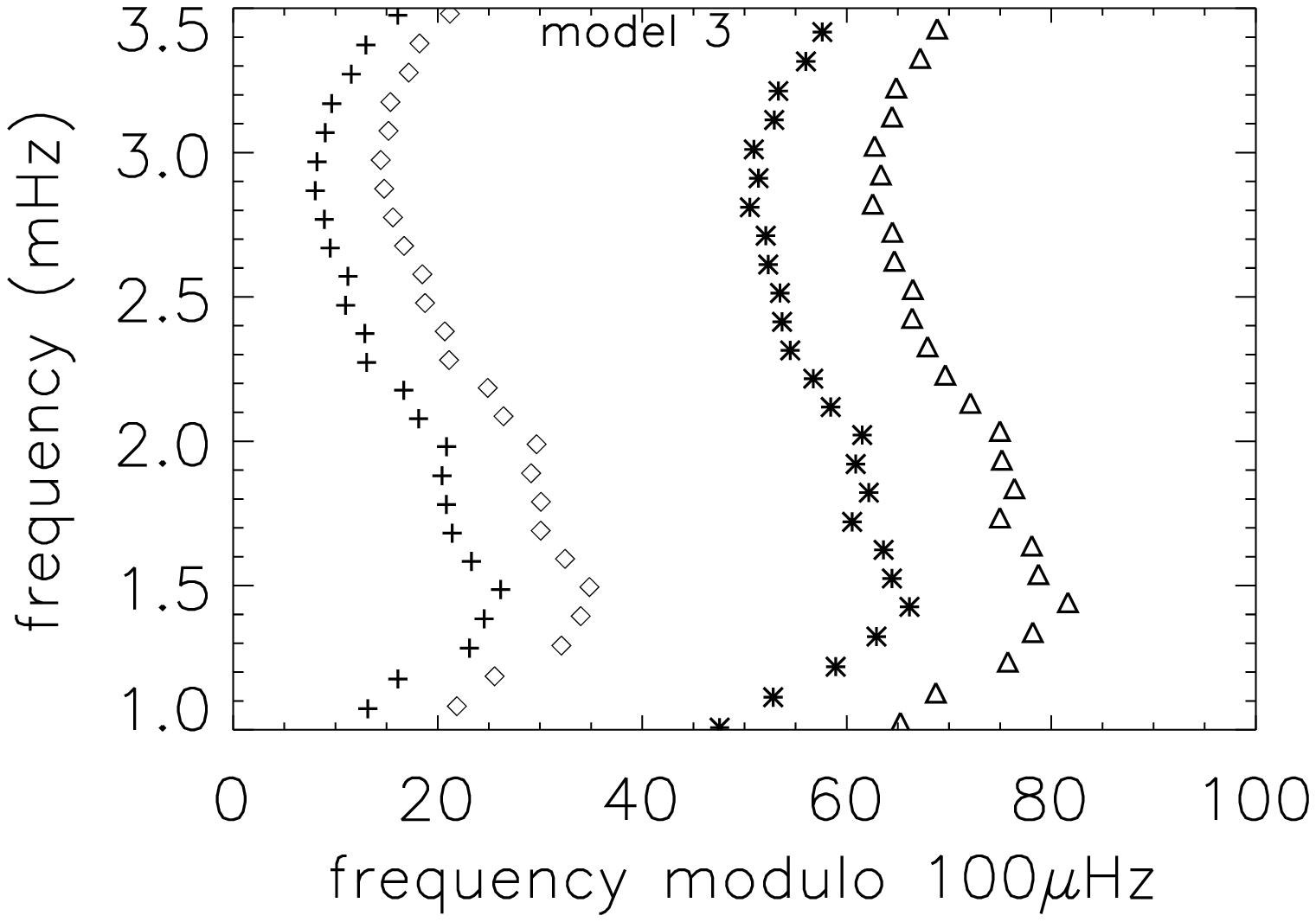}
\includegraphics[angle=0,totalheight=5.5cm,width=8cm]{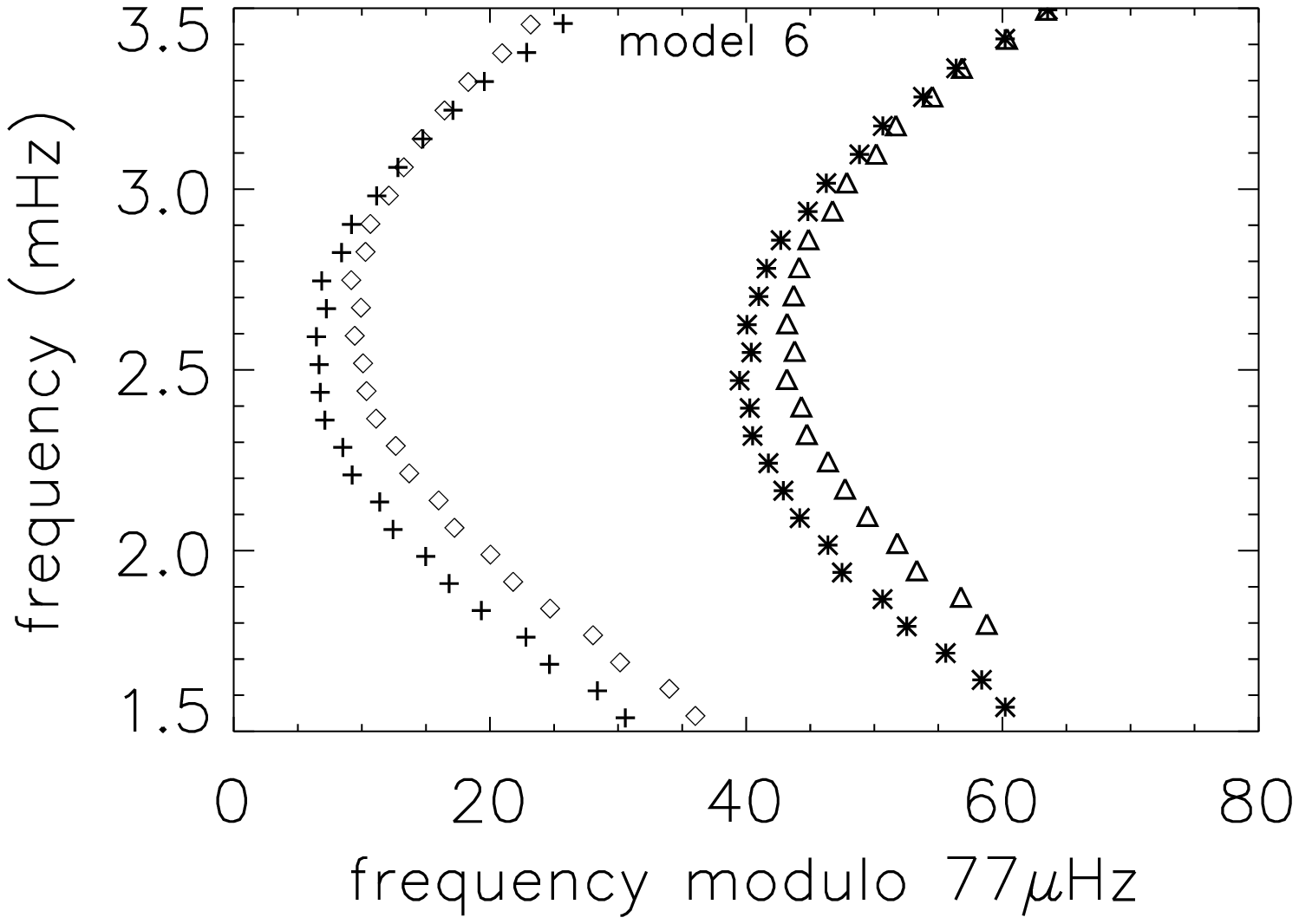}
\caption{ Echelle diagrams computed for two extreme models of HD52265, the first one with metallicity 0.27 inside the GLTR01 error box (upper panel),
 the second one with metallicity 0.19 inside the TOSKS05 error box (lower panel).\label{fig4}}
\end{figure}

\begin{figure}
\centering
\includegraphics[angle=0,totalheight=5.5cm,width=8cm]{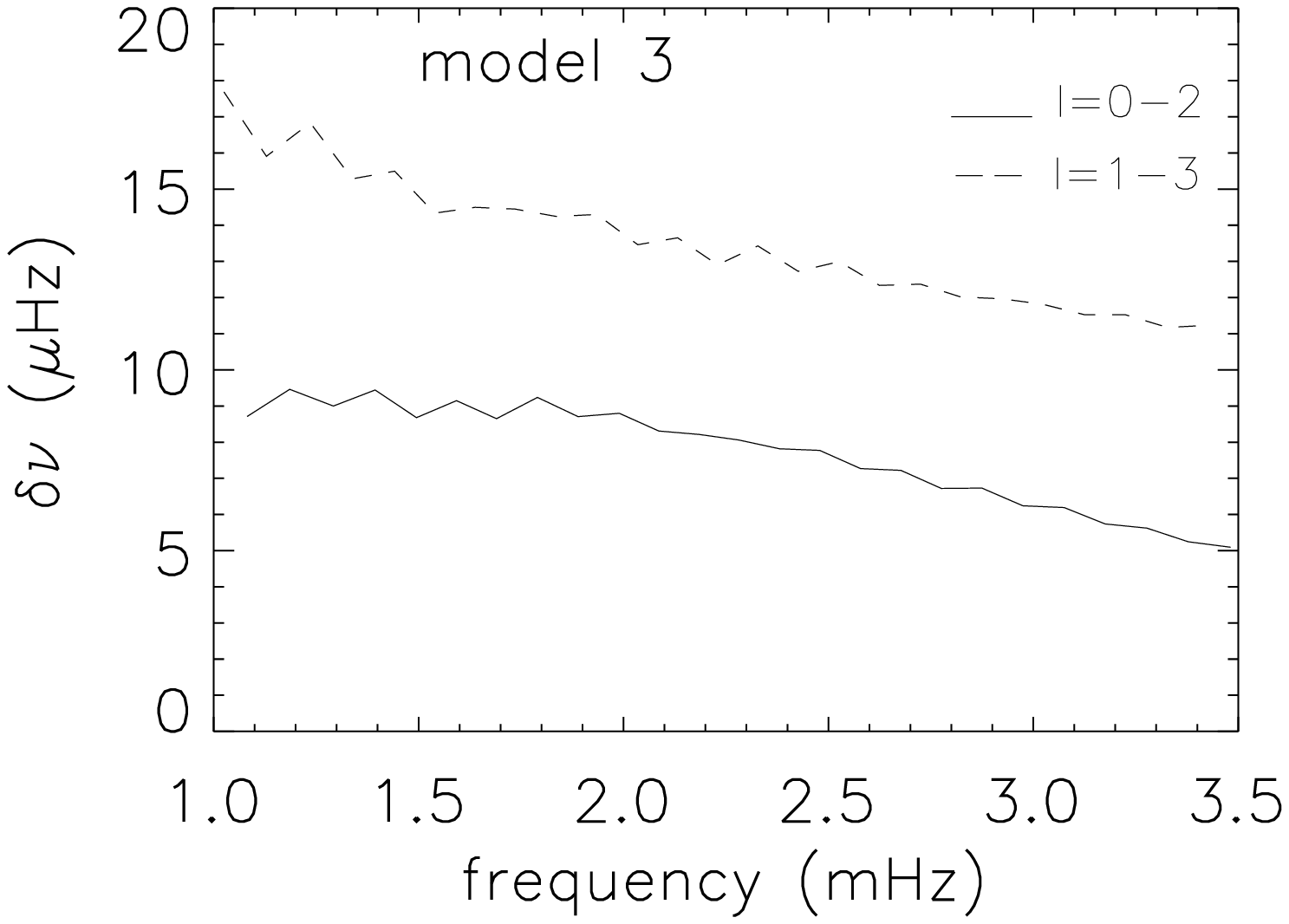}
\includegraphics[angle=0,totalheight=5.5cm,width=8cm]{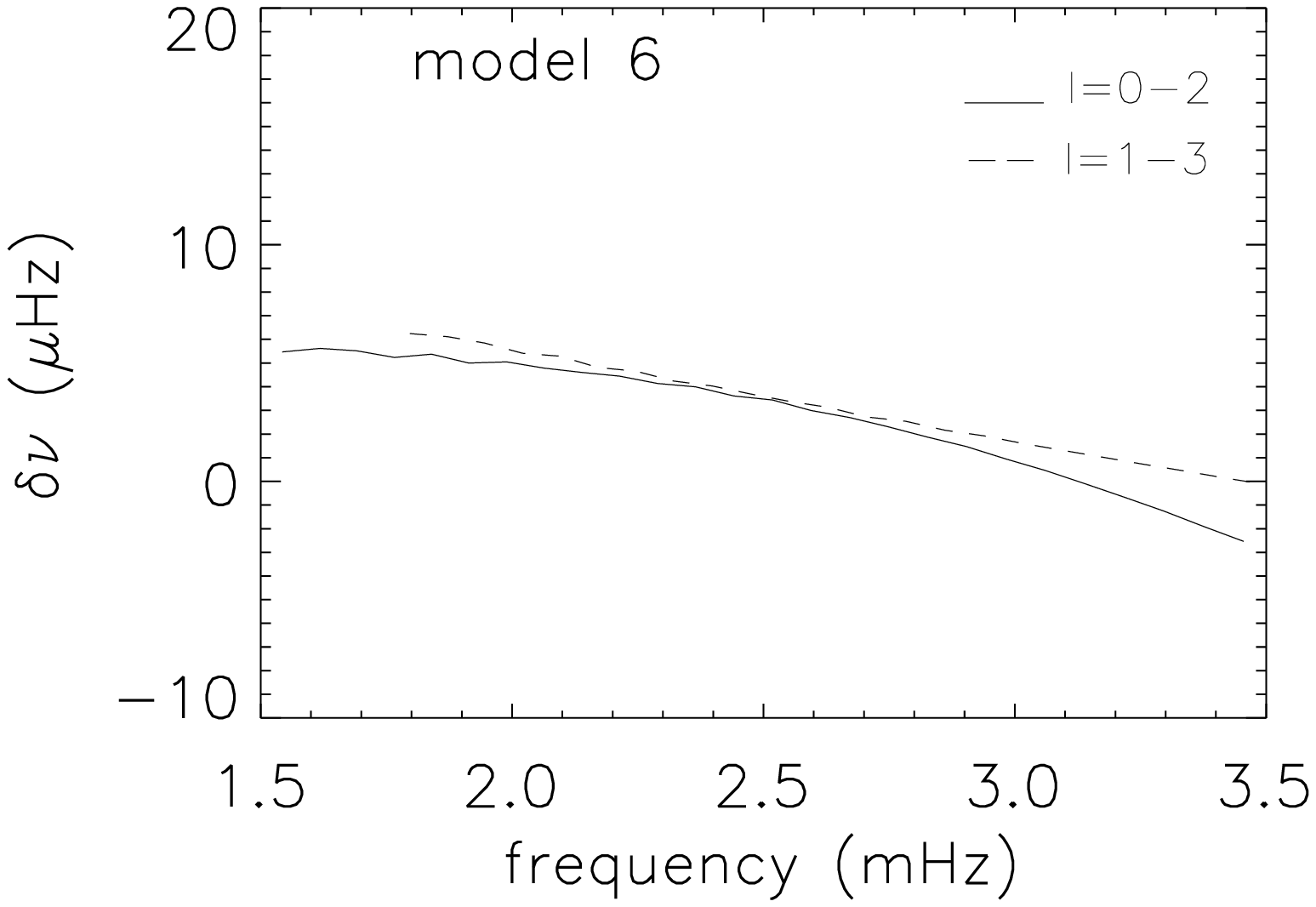}
\caption{ Small separations for the models presented in Figure 4.\label{fig5}}
\end{figure}

The echelle diagrams and small separations as obtained for two extreme models computed in the FV05 and TOSKS05 boxes are given in figure 4 and 5. They are given as examples, 
but the TOSKS05 is probably already excluded as its luminosity is too high compared to the values obtained from the Hipparcos parallax measurement. These echelle diagrams 
present interesting differences. The fact that the $\ell$=0 and $\ell$=2 lines cross at some point in the TOSKS05 case is due to the helium core, as will be discussed in 
detail in SLVV06. 
Other models, which are not presented here, lead to less spectacular differences, but in any case we are confident that COROT observations of this star will allow very 
precise tests of its internal structure.

\section{Conclusion}

The scientific community is well prepared for the future seismic observations of exoplanets-host stars which will be done with COROT, particularly the long run which will be centered on HD52265, our ``fetiche star". A large amount of data is expected, which will give the possibility of using the seismic tests with a precision never obtained before: these observations will lead to a better understanding of planetary formation and evolution.

\end{document}